\RequirePackage{fix-cm}
\documentclass[twocolumn]{svjour3}          

\smartqed  

\usepackage{latexsym}
\usepackage{subfig}
\usepackage{amssymb}
\usepackage{graphicx}
\usepackage{longtable}
\usepackage{verbatim}
\usepackage{color}
\usepackage{longtable}
\usepackage{natbib}

\newcommand{\etal}{et~al.}

\journalname{Journal of Geodesy}

\begin{document}

\title{The effects of frequency-dependent quasar variability on the celestial reference frame 
}

\titlerunning{Frequency-dependent quasar variability}        

\author{Stanislav S. Shabala         \and
  Jonathan G. Rogers \and
  Jamie N. McCallum \and
  Oleg A. Titov  \and
  Jay Blanchard \and
  James E. J. Lovell \and
  Christopher S. Watson
}

\authorrunning{Stanislav S. Shabala \etal\/} 

\institute{S. S. Shabala \at
              School of Mathematics \& Physics, University of Tasmania, Private Bag 37, Hobart, Tasmania 7001, Australia\\
              Tel.: +613-6226-8502\\
              Fax: +613-6226-2410\\
              \email{Stanislav.Shabala@utas.edu.au}           
           \and
           J. G. Rogers \at
              School of Mathematics \& Physics, University of Tasmania, Private Bag 37, Hobart, Tasmania 7001, Australia\\
              Tel.: +61-3-6226-2439\\
	     Fax: +61-3-6226-2410\\
	     \email{jgrogers@postoffice.utas.edu.au}           
           \and
           J. N. McCallum \at
              School of Mathematics \& Physics, University of Tasmania, Private Bag 37, Hobart, Tasmania 7001, Australia\\
              Tel.: +61-3-6226-7529\\
	     Fax: +61-3-6226-2410\\
	     \email{Jamie.McCallum@utas.edu.au}           
	  \and
           O. Titov \at
              Geoscience Australia, PO Box 378, Canberra, ACT 2615, Australia\\
              Tel.: +61-2-6249-9064\\
	     Fax: +61-2-6249-9929\\
	     \email{Oleg.Titov@ga.gov.au}           
           \and
           J. Blanchard \at
              School of Mathematics \& Physics, University of Tasmania, Private Bag 37, Hobart, Tasmania 7001, Australia\\
              Tel.: +61-3-6226-2449\\
	     Fax: +61-3-6226-2410\\
	     \email{jayb@utas.edu.au}           
           \and
           J. E. J. Lovell \at
              School of Mathematics \& Physics, University of Tasmania, Private Bag 37, Hobart, Tasmania 7001, Australia\\
              Tel.: +61-3-6226-7256\\
	     Fax: +61-3-6226-2410\\
	     \email{Jim.Lovell@utas.edu.au}           
           \and
           C. S. Watson \at
              School of Geography \& Environmental Studies, University of Tasmania, Private Bag 78, Hobart 7001, Australia\\
              Tel.: +61-3-6226-2489\\
	     Fax: +61-3-6226-7628\\
	     \email{Christopher.Watson@utas.edu.au}           
}

\date{Received: date / Accepted: date}

\maketitle

\begin{abstract}

We examine the relationship between source position stability and astrophysical properties of radio-loud quasars making up the International Celestial Reference Frame. Understanding this relationship is important for improving quasar selection and analysis strategies, and therefore reference frame stability. We construct flux density time series, known as light curves, for 95 of the most frequently observed ICRF2 quasars at both the 2.3 and 8.4 GHz geodetic VLBI observing bands. Because the appearance of new quasar components corresponds to an increase in quasar flux density, these light curves alert us to potential changes in source structure before they appear in VLBI images.

We test how source position stability depends on three astrophysical parameters: (1) Flux density variability at X-band; (2) Time lag between flares in S and X-bands; (3) Spectral index rms, defined as the variability in the ratio between S and X-band flux densities. We find that the time lag between S and X-band light curves provides a good indicator of position stability: sources with time lags less than $0.06$~years are significantly more stable ($>20$ percent  improvement in weighted rms) than sources with larger time lags. A similar improvement is obtained by observing sources with low ($<0.12$) spectral index variability. On the other hand, there is no strong dependence of source position stability on flux density variability in a single frequency band. These findings can be understood by interpreting the time lag between S and X-band light curves as a measure of the size of the source structure. Monitoring of source flux density at multiple frequencies therefore appears to provide a useful probe of quasar structure on scales important to geodesy.

The observed astrometric position of the brightest quasar component (the core) is known to depend on observing frequency. We show how multi-frequency flux density monitoring may allow the dependence on frequency of the relative core positions along the jet to be elucidated. Knowledge of the position-frequency relation has important implications for current and future geodetic VLBI programs, as well as the alignment between the radio and optical celestial reference frames.

\end{abstract}

\section{Introduction}
\label{sec:intro}

High-precision studies of the Earth system, as well as spacecraft tracking, place stringent requirements on space geodetic techniques, and in particular on the realization of accurate and stable Celestial \citep[CRF; ][]{MaEA09} and Terrestrial \citep[TRF; ][]{AltamimiEA07} Reference Frames. For example, the next generation VLBI system, the VLBI Geodetic Observing System (VGOS), aims to measure station positions to 1~mm accuracy, and station velocities to 0.1~mm~year$^{-1}$ \citep{NiellEA06,PetrachenkoEA09}. This is an order of magnitude better than current measurements \citep{SchuhBehrend12,LovellEA13}. Clearly, many systematic and stochastic sources of error will need to be eliminated, or at least substantially mitigated to reach this goal.

\citet{PetrachenkoEA09} performed a simulation study that considered three sources of stochastic noise: wet troposphere delay, station clocks and measurement error, and concluded that wet troposphere is the most important of these. Fast ($\sim 30$~seconds) source-switching times were predicted to reduce station position root-mean-square (rms) error to the millimetre level. Such rapid switching is possible with the new class of small, fast-slewing 12-metre antennas currently being constructed around the world. Further gains are being made by improving the International VLBI Service for Geodesy and Astrometry (IVS) network geometry through addition of more stations in the southern hemisphere, such as the AuScope array in Australia \citep{LovellEA13} and the Warkworth antenna in New Zealand \citep{WestonEA13}. 

A dramatic reduction in wet troposphere noise means other potential sources of error must be considered more carefully. One of these is the structure of radio-loud quasars making up the International Celestial Reference Frame (ICRF2) \citep{MaEA09}. In a seminal work, \citet{Charlot90} developed the formalism for estimating the effects of quasar structure on VLBI group delays. VLBI imaging allows the structure index \citep[SI; ][]{FeyCharlot97} of an individual source to be calculated. This quantity is related to the logarithm of the median time delay due to quasar structure observed with all terrestrial baselines. Following the work of \citet{FeyEA96}, \citet{FeyCharlot97,FeyCharlot00} and \citet{OjhaEA04}, \citet{MaEA09} tabulated the median structure index values for more than 700 ICRF2 quasars, by combining multi-epoch VLBI images of these sources. These authors recommended that quasars with median SI greater than 3 (corresponding to an additional group delay of 10~picoseconds, or 3 mm in light travel) should be considered unsuitable for geodesy.

VLBI imaging at the geodetic frequency of 8.4~GHz typically probes quasar structure on scales of 0.1~mas or larger. For comparison, a quasar consisting of two equal brightness components separated by 60~$\mu$as can in principle result in a position shift of 1.5 mm on a 10,000~km baseline. Observations of multiple quasars will reduce this shift in estimated station positions, however the effects of quasar structure will still be apparent in the scatter in determined station positions, as recently shown in simulations by \citet{ShabalaEA14b}. This means that even quasars that have no apparent structure in VLBI images could potentially affect geodetic measurements\footnote{This sub-resolution structure will be reflected in visibility amplitudes and phases of VLBI measurements, suggesting the presence of substructure on scales smaller than the synthesized beam.}. We refer to this as the {\it sub-resolution} issue.

The phenomenon of Interstellar Scintillation (ISS) offers an indirect way of probing much smaller angular scales than those available through VLBI imaging. Scintillation is the process of rapid (hours to days) variability in the apparent flux density of a radio source. For a source to exhibit ISS, a substantial fraction of its flux density must come from angular scales not exceeding $10-50$~$\mu$as \citep{LovellEA08}, the typical angular size of the clump of ionised gas crossing the observer's line of sight and causing the scintillation. VLBI observations show that scintillating sources show less structure on milliarcsecond scales than non-scintillating quasars \citep{OjhaEA04,OjhaEA06}. In a recent paper, \citet{SchaapEA13} showed that scintillators have more stable positions than non-scintillating sources, and recommend that scintillation be used as a criterion for selection of new radio sources in future reference frame realizations. Observations of scintillation at multiple frequencies allows quasar structure to be quantified on scales of 10~$\mu$as \citep{TurnerEA12,MacquartEA13}, thus potentially probing the sub-resolution regime.

The above analysis ignores the temporal evolution of quasars. Detailed studies of jet kinematics performed by the MOJAVE\footnote{www.physics.purdue.edu/astro/mojave} team \citep{ListerEA09b} show that many flat-spectrum radio quasars undergo regular outbursts. In VLBI images, this is manifested through core brightening, followed by a separation of the ejected component from the core with proper motion of $0.05-0.5$~mas~year$^{-1}$ \citep{KellermanEA04,ListerEA09b}. These changes on VLBI scales are correlated with source {\it flaring} in total source flux density, where the flux density increases to a peak value which is often several times higher than the pre-flare flux density, and then falls again \citep[e.g.][]{TerasrantaEA05,KudryavtsevaEA11}. Importantly, flare characteristics often change with frequency, as expected from models of self-absorbed synchrotron jets \citep{MarscherGear85,Hirotani05,KudryavtsevaEA11,SSG12}. 

Motivated by this link between quasar variability and structure, in the present contribution we investigate the effects of quasar flux density variability on geodetic solutions. In Section~\ref{sec:data} we describe the data used in this work. Section~\ref{sec:analysis} outlines our analysis procedure, and results are presented in Section~\ref{sec:results}. We discuss the implications of our findings in Section~\ref{sec:discussion}, and conclude in Section~\ref{sec:conclusions}.
  
\section{Data}
\label{sec:data}

\subsection{Light curves}
\label{sec:fluxData}

\begin{figure}
  	\includegraphics[width=0.35\textwidth,clip,angle=270]{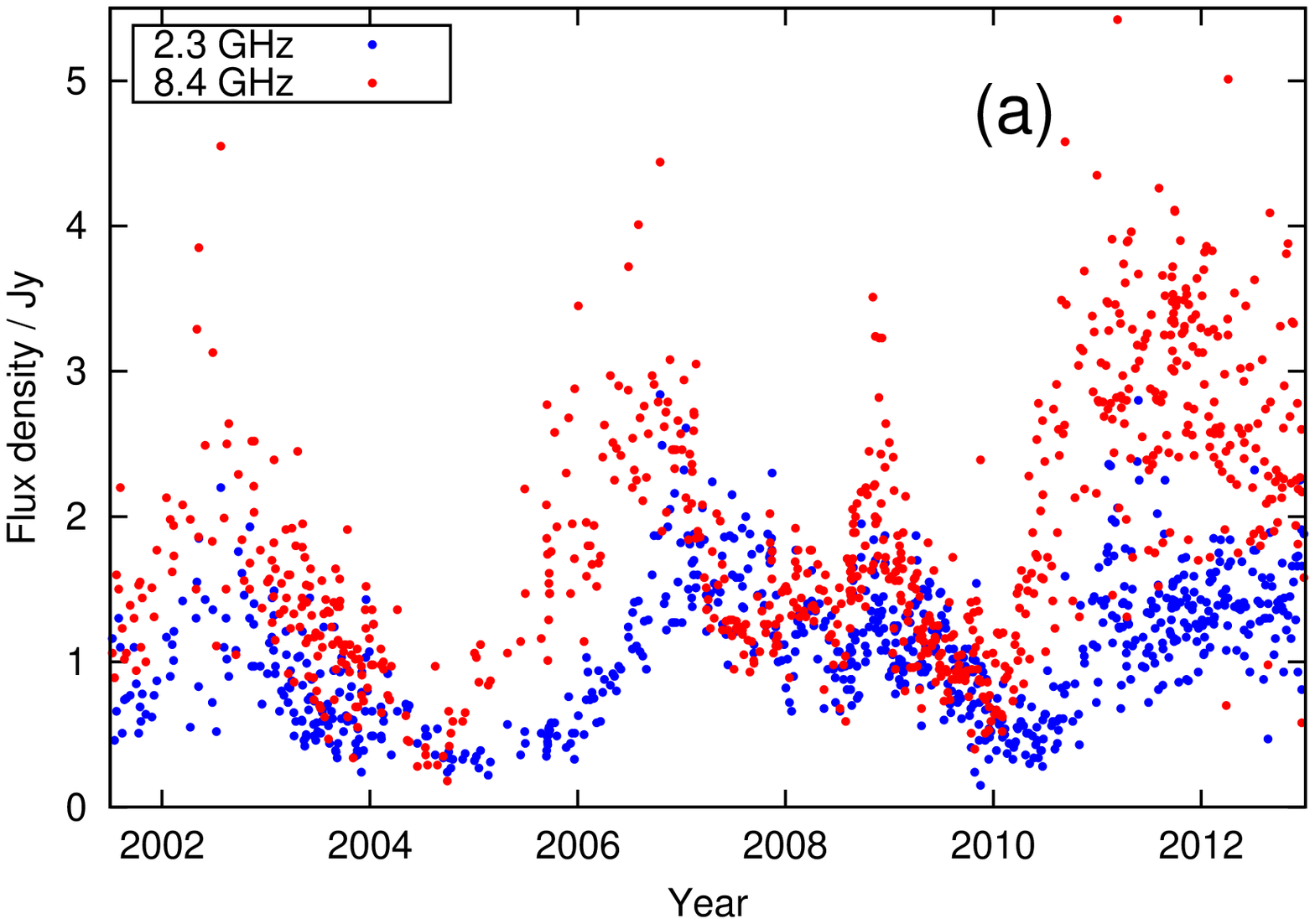} 
	\includegraphics[width=0.35\textwidth,clip,angle=270]{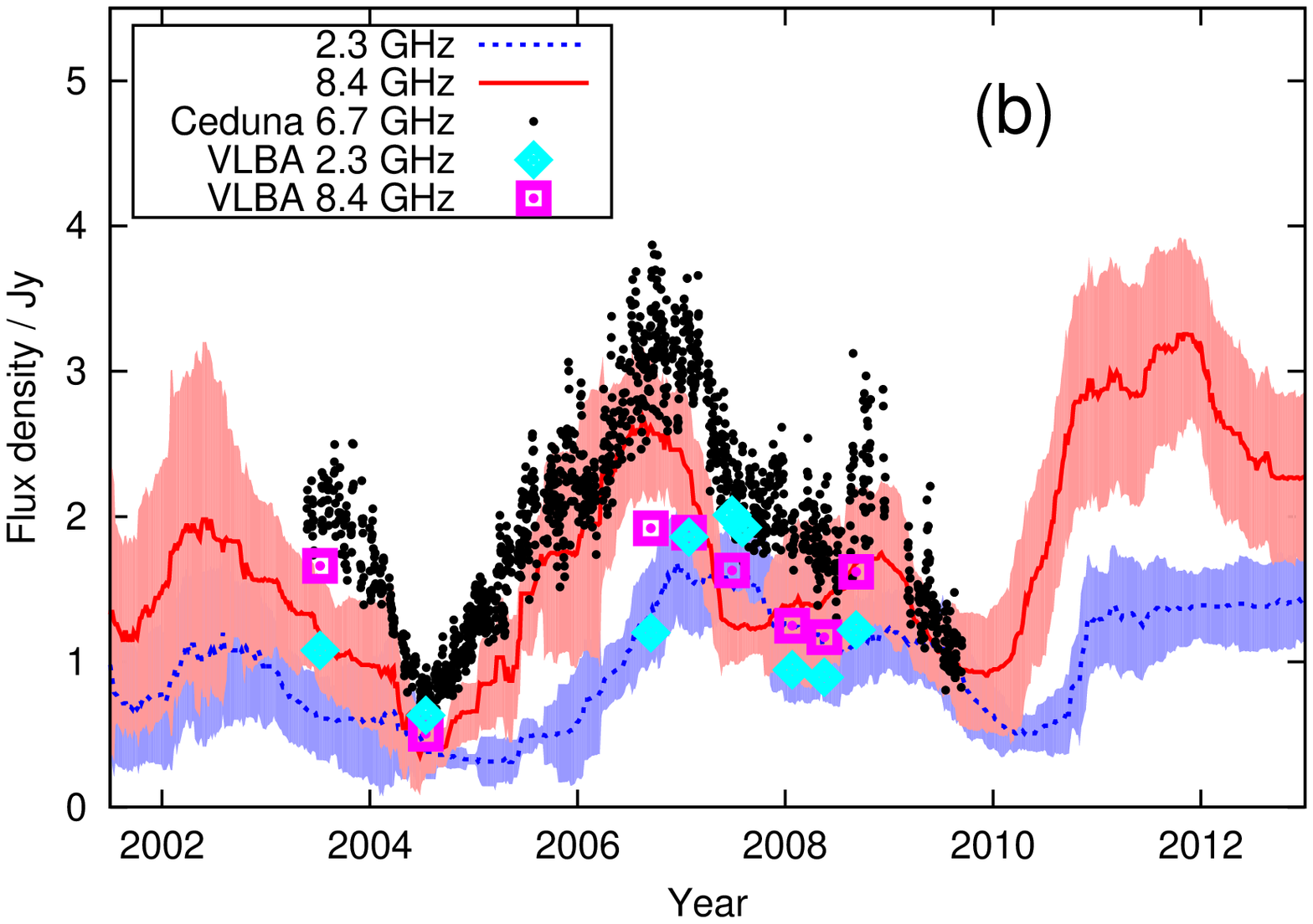} 
\caption{Flux density variability of quasar PKS~B1144--379 at X and S-bands. {\it (a)}: Raw data from IVS observations. {\it (b)}: Blue and red shaded regions show the raw IVS data smoothed with a 200-day window, with 1$\sigma$ uncertainties. Also shown are single-dish 6.7~GHz data from the UTas Ceduna observatory flux density monitoring (black points) and VLBI data from USNO at S-band (cyan) and X-band (purple). The averaged IVS data set agrees well with these well-calibrated flux density measurements. We note that 6.7~GHz is close enough in frequency to X-band (8.4~GHz) to expect qualitatively similar behaviour in the two light curves, albeit with different flux density scales.}
\label{fig:1144_lc}
\end{figure}

In this work we use the 2.3~GHz (S-band) and 8.4~GHz (X-band) flux density time series, known as {\it light curves}. These are constructed using standard IVS products \citep{SchuhBehrend12}. Flux densities of individual sources have been estimated as part of the correlation process since 2000, and are available on the Goddard Space Flight Center (GSFC) server\footnote{lupus.gsfc.nasa.gov/sess/sessions/cumulative/source\_perf/}. Because accurate flux density calibration is not a key requirement in geodetic VLBI, these values must be validated against well-calibrated data. In Figure~\ref{fig:1144_lc} we show a comparison between the IVS flux densities and two sets of astronomical measurements for the IVS source PKS~B1144--379. The top panel shows the raw IVS data at S and X-bands. In the bottom panel, the red and blue shaded regions show the same data smoothed with a 200-day window function. Accurate measurements of VLBI core fluxes from the imaging data of the US Naval Observatory (USNO)\footnote{rorf.usno.navy.mil/rrfid.shtml} at the same frequencies are shown as large diamonds or squares; these are broadly consistent with the averaged raw data.

Because VLBI imaging is typically performed at much lower cadence than either IVS or single-dish flux density monitoring observations, we also use data from the University of Tasmania's COSMIC project \citep{McCullochEA05,CarterEA09} for validation. This project uses the 30-metre antenna at Ceduna, South Australia, to monitor flux densities of many IVS quasars at 6.7~GHz. This frequency is close enough to the IVS X-band ($8.2-8.95$~GHz) to be useful for validating the X-band data (see Section~\ref{sec:coreShift}). Although at any given time the flux densities will be different at the two frequencies due to a non-zero spectral index and also the different physical regions of the quasar sampled by the Ceduna single dish and the (much higher resolution) IVS network, the general trends associated with individual flares should be similar because geodetic quasars are selected for their compactness. This is confirmed by the observed correlation between time series represented by black dots and red lines in Figure~\ref{fig:1144_lc}{\it b}. We performed these validation tests for four representative sources for which we had sufficient data at 8.4 and 6.7~GHz, and found that the averaged IVS flux density data agreed well with well-calibrated single-dish and VLBI flux density measurements.

\subsection{Position time series}
\label{sec:positionData}

Source position time series were obtained from IVS observations using the OCCAM geodetic analysis software package. The data were processed using the least squares collocation technique \citep{TitovEA04}, adopting the Vienna Mapping Function \citep[VMF1;][]{BoehmEA06_vmf1} and corrections for precession, nutation, solid tide, ocean loading, pole tide including the mean pole model and thermal expansion consistent with the IERS Conventions 2010 \citep{PetitLuzum10}. We do not, however, implement non-tidal atmospheric loading. By utilising covariance matrices, the least squares collocation technique takes into account correlations between the observables within a given 24-hour session. This allows stochastic parameters to be estimated on a scan-by-scan basis (rather than the hourly timescales associated with the conventional linear least squares method), and has the effect of reducing tropospheric noise. Coordinates of well-established stations were used to impose the No-Net-Rotation (NNR) and No-Net-Translation (NNT) constraints. A No-Net-Rotation condition on defining ICRF2 sources was imposed to fix the orientation of the Celestial Reference Frame. Station coordinates, Earth Orientation Parameters and source positions were estimated simultaneously once per session for all sources. In total, 67,070 positions of 2038 sources were estimated in 1294 geodetic sessions over a 10-year period from 2000--2009.

\section{Analysis}
\label{sec:analysis}

The major aim of this work is to test how source position stability depends on physical properties of the source (as inferred from various astrophysical indicators). In this section, we construct metrics to quantify both position stability and astrophysical properties of the ICRF2 quasars. 

\subsection{Position stability}

\begin{figure}
  	\includegraphics[width=0.4\textwidth,clip,angle=270]{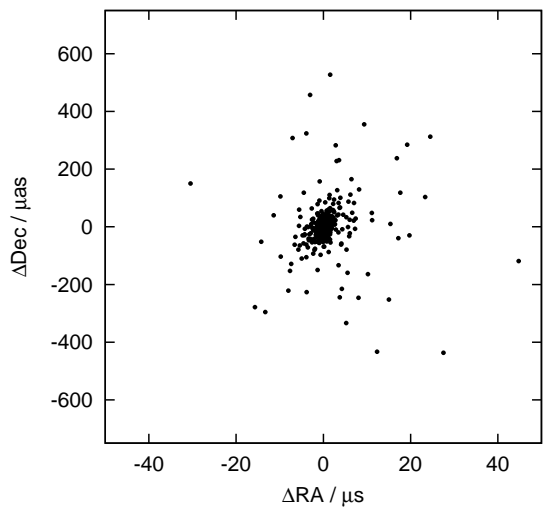} 
\caption{Source positions estimated using OCCAM for source PKS~B1144--379. The plotted position offsets are relative to the ICRF2 source coordinates. The median offset in position from the origin is 17~$\mu$as.}
\label{fig:1144_pos}
\end{figure}

The metric we adopt to define solution quality is source position repeatability. If all quasars were point-like, there was no thermal noise, and the modelling of Earth dynamics and kinematics as well as signal propagation effects were perfectly accurate, the same source positions would be recovered every time. In reality, all these effects conspire to introduce scatter in estimated source positions, as shown in Figure~\ref{fig:1144_pos}.

The accuracy of source position estimates is highly dependent on the quality of atmospheric solutions; these in turn are largely determined by network geometry. On the other hand, the precision of source position estimates (i.e. the size of the formal error bars) for a typical geodetic schedule depends on the number of observations of the source in a given session. For our full sample, the median formal uncertainty is $17.9$~$\mu$as. Although in many cases the formal uncertainties in source positions are larger than the position offsets (for example, the median position offset from the catalogued coordinates for source PKS~B1144--379 shown in Figure~\ref{fig:1144_pos} is 17~$\mu$as), these are not entirely representative of true uncertainties and we choose to ignore them in our analysis. We note that all subsamples we consider below have very similar distributions of formal uncertainties, and our results should therefore be immune to this aspect of the analysis.

The presence of source structure (i.e. quasars not being point-like) will also induce scatter in source positions. This is because the quasar brightness distribution will contribute an additional term to interferometer visibility phase as measured on any given baseline. The measured VLBI group delay is just the slope of phase across the bandpass at X-band. Because the sign and magnitude of this phase term depend on the {\it projected} quasar structure as viewed by a given baseline, quasars with significant structure will register very different contributions to the group delay when observed at different times and with different pairs of antennas. Therefore, quasars with significant structure are expected to yield a large scatter in source positions when observed under different circumstances (e.g. using different schedules). This source structure contribution to scatter in estimated source positions is in addition to the scatter due to network geometry, which should be independent of quasar physics.

Motivated by these considerations, we use as a metric of source position repeatability the fraction of source positions that are $>200$~$\mu$as away from the mean source position. In other words, for each position estimate we calculate the quantity

\begin{equation}
\Delta = \left[ \left( \Delta {\rm RA} \cos ({\rm Dec}) \right)^2 + \left( \Delta {\rm Dec} \right)^2 \right]^{1/2} \nonumber
\end{equation}
and then for each source calculate the metric

\begin{equation}
f_{200} = N \left( \Delta > 200\mu{\rm as} \right) / N_{\rm total}
\label{eqn:f20}
\end{equation}
where $N_{\rm total}$ is the number of total position measurements for that source. In this definition, position scatter was preferred over individual coordinate scatter because the individual RA and Dec coordinates estimated by OCCAM were found to be not very reliable, with most of the offset usually being absorbed in one coordinate \citep{SchaapEA13}. The radius of $200$~$\mu$as corresponds to 3.5 times the ICRF2 noise floor of 40~$\mu$as in each coordinate \citep{MaEA09}. We note that we repeated our analysis below for cutoff radii of 100, 400 and 600~$\mu$as, and found qualitatively similar results. In what follows, we therefore adopt $f_{200}$ as a measure of source position stability.

\subsection{Variability metrics}
\label{sec:varMetrics}


Detailed studies of radio-loud quasars \citep[e.g.][]{KellermanEA04,ListerEA09b} have shown that ejection of milliarcsecond-scale components seen in VLBI images is accompanied by an overall source brightening. Physically, new jet components are created by temporary changes in the regions near the black hole responsible for the observed radio emission. These new jet components contribute to the increase in source brightness, and are initially coincident with the bright central component (known as the ``core''; see Section~\ref{sec:interpretation}) before gradually moving outwards at speeds of up to $0.5$~mas~year$^{-1}$ \citep{ListerEA09b}. VLBI imaging at X-band can only clearly resolve the jet components when they are $>100$~$\mu$as from the core \citep{KovalevEA08,SokolovskyEA11}. On the other hand, flux density monitoring (with VLBI or using a single dish) can identify ejection of a new component (by detecting the start of a flare) long before it is seen in VLBI images\footnote{For typical flat-spectrum quasar proper motion of 0.2~mas~year$^{-1}$ \citep{ListerEA09b}, VLBI imaging at X-band will only resolve the new jet component 6 months after the beginning of the flare.}. Of course, the ideal data set would consist of VLBI imaging performed at the cadence of present-day flux density monitoring (at least once a week). Such VLBI images could be used to construct the integrated light curves for the whole source, similar to the ones presented here; as well as providing spatial information on where the emission is coming from (e.g. inside or outside the core) and how much structure the source has on 100~$\mu$as scales. While such frequent imaging may be possible in the future with the next-generation VLBI observing system \citep{PetrachenkoEA09}, it is not available at present.

In the absence of frequent VLBI imaging, the two-frequency quasar light curves shown in Figure~\ref{fig:1144_lc} can be used to construct several useful astrophysical metrics potentially connected to source structure.

\subsubsection{Modulation index}


For a quasar exhibiting structure, both separation of the components and their relative brightness are important in calculating the source structure contribution to the observed group delay \citep[e.g.][]{Charlot90}. Flare amplitude is an indicator of the relative brightness of the core and jet components: the higher the amplitude, the brighter the jet component relative to the core. We quantify flare amplitude via a {\it modulation index},

\begin{equation}
	\mu = \sigma_F / \bar{F}
\label{eqn:mi}
\end{equation}
where $\bar{F}$ is the mean flux density at X-band, and $\sigma_F = \left[ \sum_{i=1}^N (F_i - \bar{F} )^2 / N \right]^{1/2}$ the root mean square (rms) about this mean value. We choose to use X-band flux densities because this is the frequency at which source positions are estimated; in addition, radio source variability typically increases with frequency \citep{TerasrantaEA05,KudryavtsevaEA11}. High values of $\mu$ indicate launching of bright new jet components, and therefore are an indicator of potential source structure evolution.

\subsubsection{Time lags}
\label{sec:timeLags}

Figure~\ref{fig:1144_lc} shows that quasar variability for PKS~B1144--379 is qualitatively similar at both S and X-bands, however there is a clear {\it time lag} after year 2004, with the X-band light curve leading. Inspection of light curves revealed similar behaviour for most sources in our sample. This echoes the findings of \citet{KudryavtsevaEA11} for the quasar 3C345. These authors interpreted the observed time lag between light curves at various radio frequencies as a manifestation of the so-called ``core shift'', a phenomenon resulting from self-absorption of the synchrotron jets in flat-spectrum quasars. We discuss the core shift and its relation to the time lags in Section~\ref{sec:interpretation}. Due to this relation, it is thus of interest to use time lags as another metric of the physical properties of the quasars.

In order to calculate time lags for individual sources, we applied the Discrete Correlation Function (DCF) to IVS flux density data smoothed with a 200-day window function. The DCF \citep{EdelsonKrolik88} is a generalised cross-correlation function that is robust to uneven data sampling. For reliability, only sources with at least 300 flux density measurements over the 10-year observing period were used in our analysis; this yielded a sample of 95 sources. Inspection of Figure~\ref{fig:1144_lc} shows that one must take care when computing the DCF: for a given source, the time lag between X and S-band light curves can change appreciably between flares. This can be understood by invoking the result of \citet{SSG12} that more powerful jet outbursts lead to larger core shifts (and therefore time lags). Each flare corresponds to an independent outburst, and therefore the S/X time lags will be different for individual flares.

\begin{figure}
  	\includegraphics[width=0.35\textwidth,clip,angle=270]{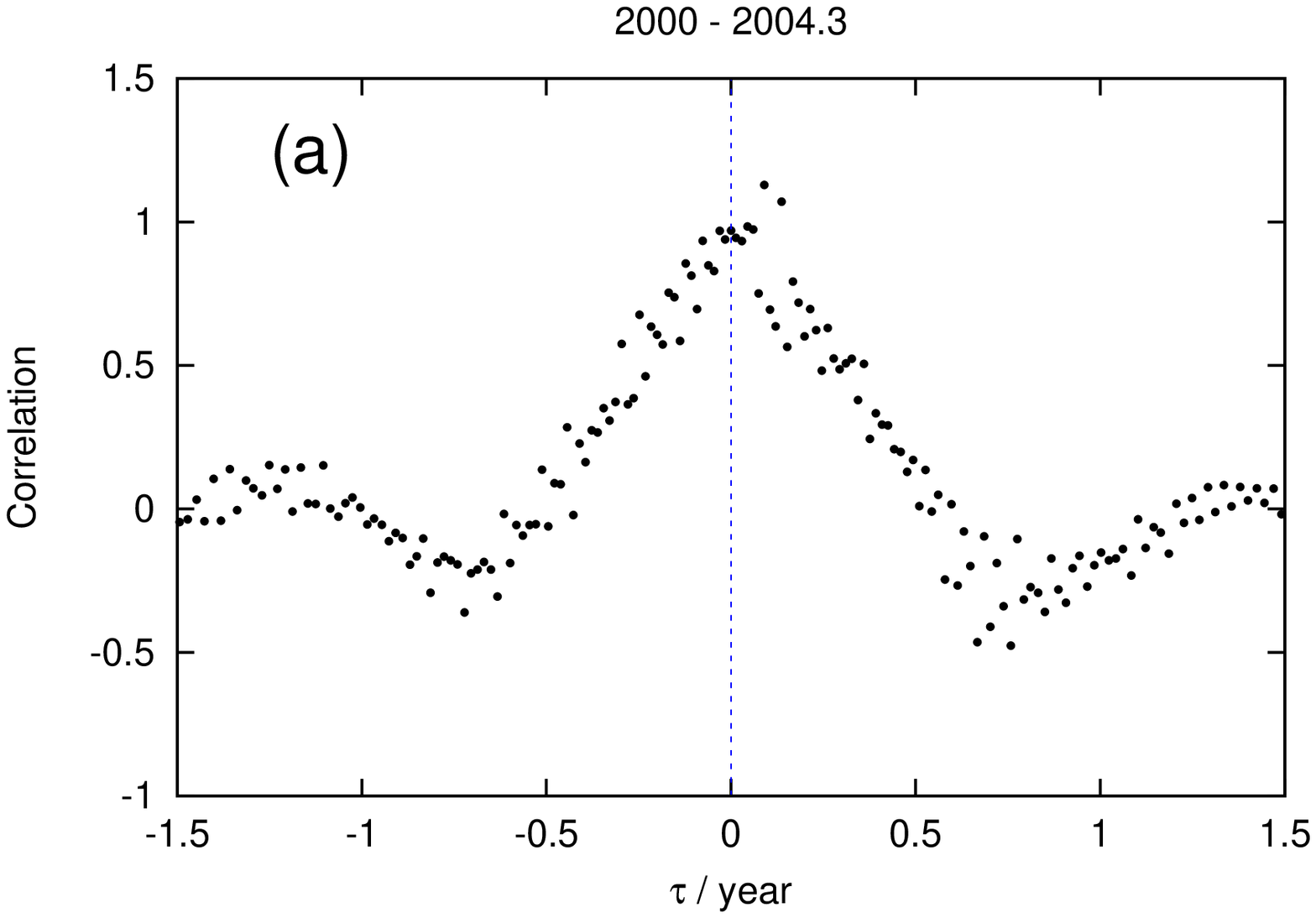} 
	\includegraphics[width=0.35\textwidth,clip,angle=270]{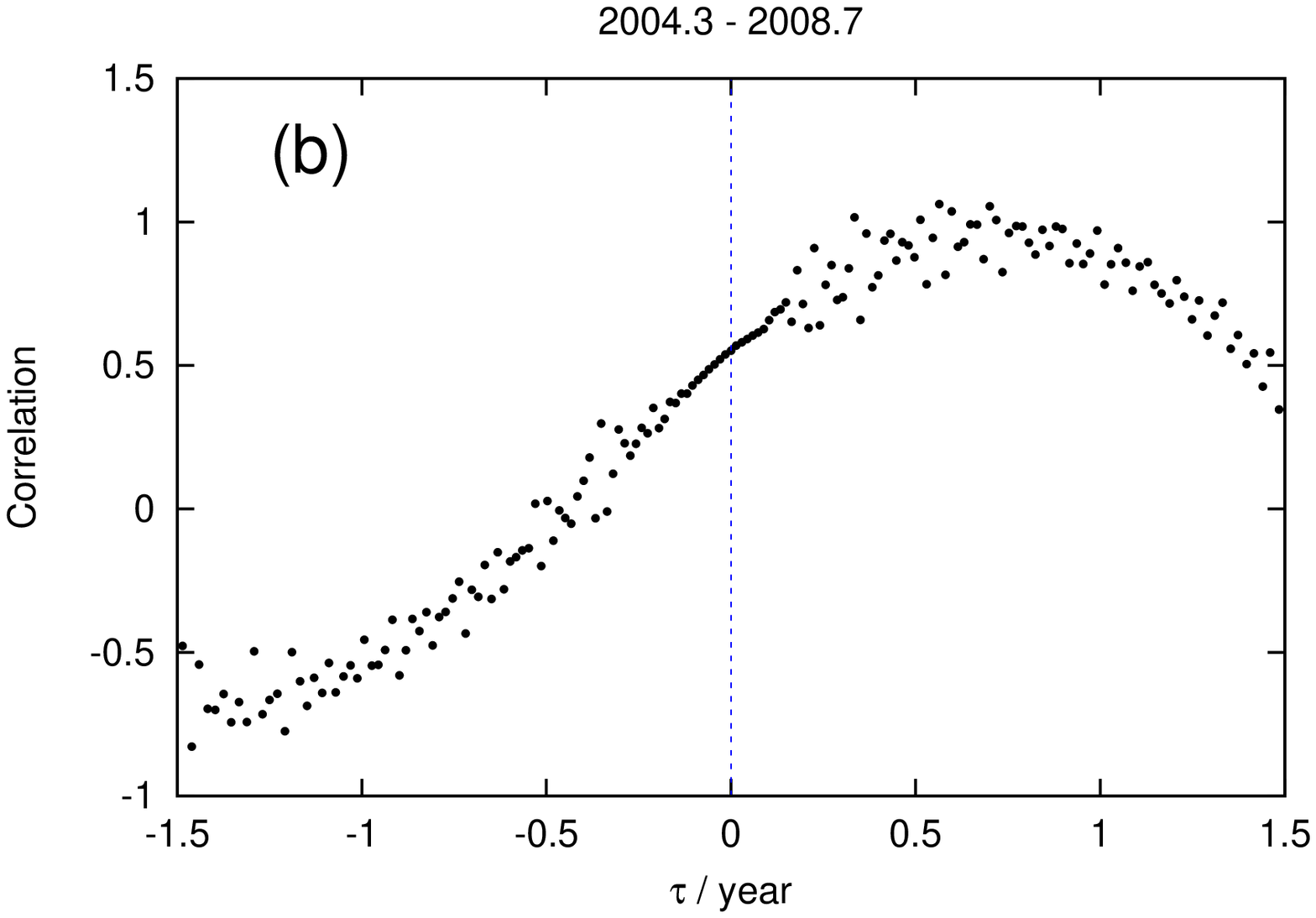} 
	\includegraphics[width=0.35\textwidth,clip,angle=270]{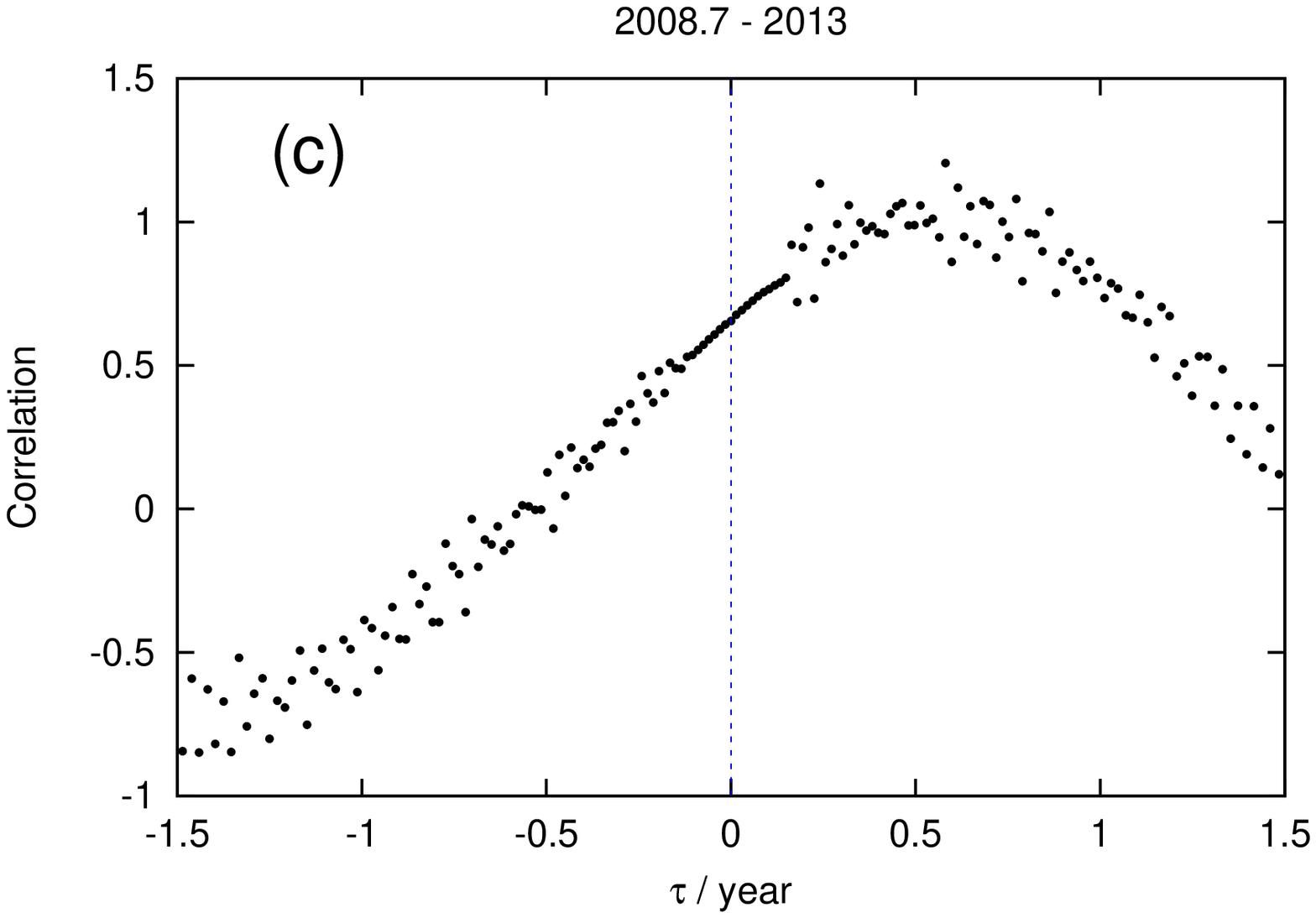}
\caption{Discrete Correlation Functions computed using the S and X-band light curves for representative flares in the quasar PKS~B1144--379. Because each flare has its own characteristic time lag, the 13 years of data are split into three segments. The middle segment contains two flares blended together.}
\label{fig:1144_DCFs}
\end{figure}

In Figure~\ref{fig:1144_DCFs} we split the flux density time series for the source PKS~B1144--379 into three equal length segments (one for each de-blended flare), and show the DCFs for these. The peak in the DCF corresponds to the best-fit time lag between S and X-band light curves. Our DCF analysis clearly recovers the (almost) zero time lag for the first segment, and $\tau \sim 0.6$ years for the last two flares. In what follows, we divide all light curves into equal length segments such that the time lag does not change appreciably in any given segment; a 10-year time series is therefore divided into two to four segments for each source, each segment typically containing a single flare.

\subsubsection{Spectral index variability}

\begin{figure}
  	\includegraphics[width=0.35\textwidth,clip,angle=270]{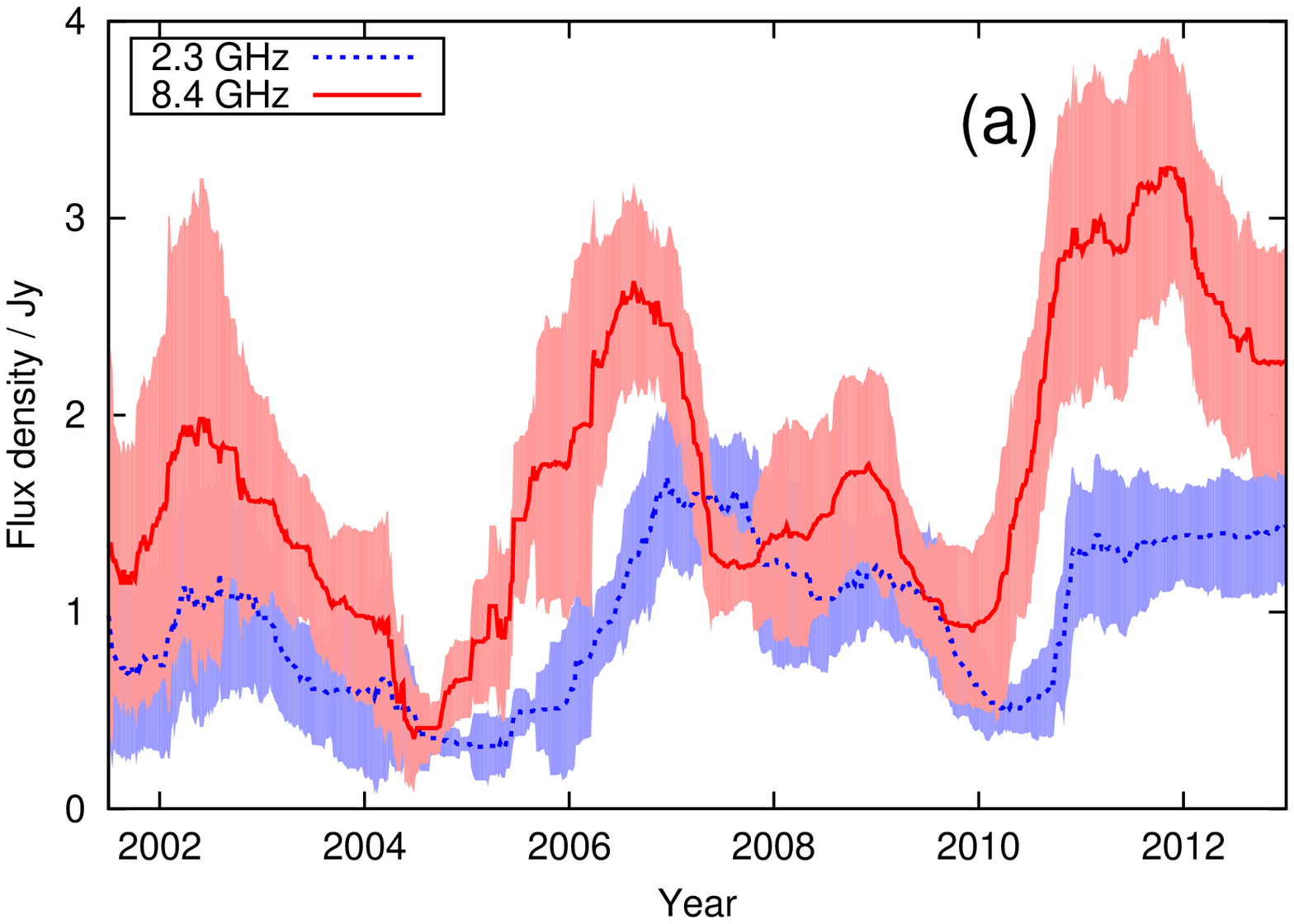} 
	\includegraphics[width=0.35\textwidth,clip,angle=270]{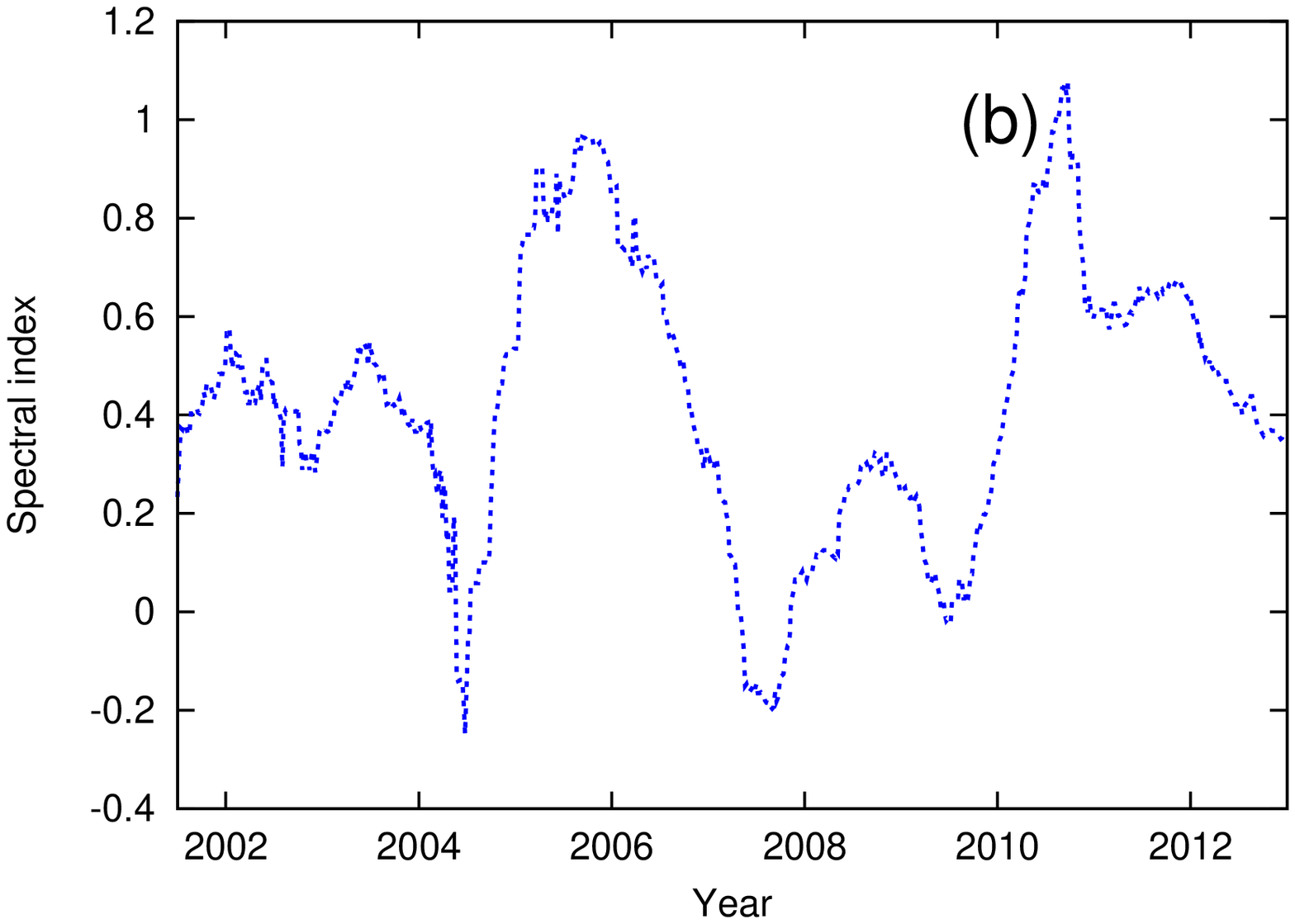} 
\caption{{\it (a)}: Flux density variability of quasar PKS~1144--379 at X and S-bands. The data are from IVS observations, averaged over 200 days. {\it (b)}: Two-point spectral index. Before 2004 the variability at X and S bands occurs ``in phase'', resulting in little change in the source spectral index. Later flares show a clear time lag between the two bands, with X-band emission leading; this is reflected in the large variability in the spectral index.}
\label{fig:1144_spInd}
\end{figure}

The two metrics described above may be insufficient: modulation index quantifies source variability, but does not contain any spatial information; on the other hand, time lags may provide spatial information (through the relation with the core shift) but do not quantify source variability. We therefore construct a final metric which depends on both the magnitude of source variability and the time lag between the S and X-band light curves. We define the {\it two-point spectral index} as
\begin{equation}
	\alpha = \log (F_X / F_S) /  \log (8.4/2.3)
\label{eqn:si}
\end{equation}
This is the standard astronomical definition of the spectral index, and is simply the slope of flux density against frequency in log-log space. As we show in Figure~\ref{fig:1144_spInd}, the two-point spectral index is sensitive to changes in the quasar light curves: when variability in both bands occurs in phase and is similar in amplitude, the spectral index remains constant; however the spectral index begins to vary significantly as soon as a time lag appears.

We therefore use the spectral index rms as the third and final metric of source variability,

\begin{equation}
	\sigma_{\alpha} = \left[ \sum_{i=1}^N \frac{(\alpha_i - \bar{\alpha})^2} {N} \right]^{1/2}
\label{eqn:si_rms}
\end{equation}

\subsection{The sample}

Our final sample is constructed as follows. Starting with all observed ICRF2 quasars, we only retain the 95 quasars for which we have at least 300 flux density measurements; this is required for robust construction of the averaged light curves. The light curves are split into a total of 330 segments, typically 3--4 equal length segments for each source. For each segment, we determine the time lag between S and X-band light curves using the DCF. Only segments with robust time lags (cross-correlation peak $>0.7$) are retained. We also only keep segments with more than 50 position measurements in order to minimise Poisson errors in our analysis. This yields a final sample of 84 segments with time lags, modulation indices, spectral index rms and position measurements.

\section{Results}
\label{sec:results}

\begin{figure*}
  	\includegraphics[width=0.35\textwidth,clip,angle=270]{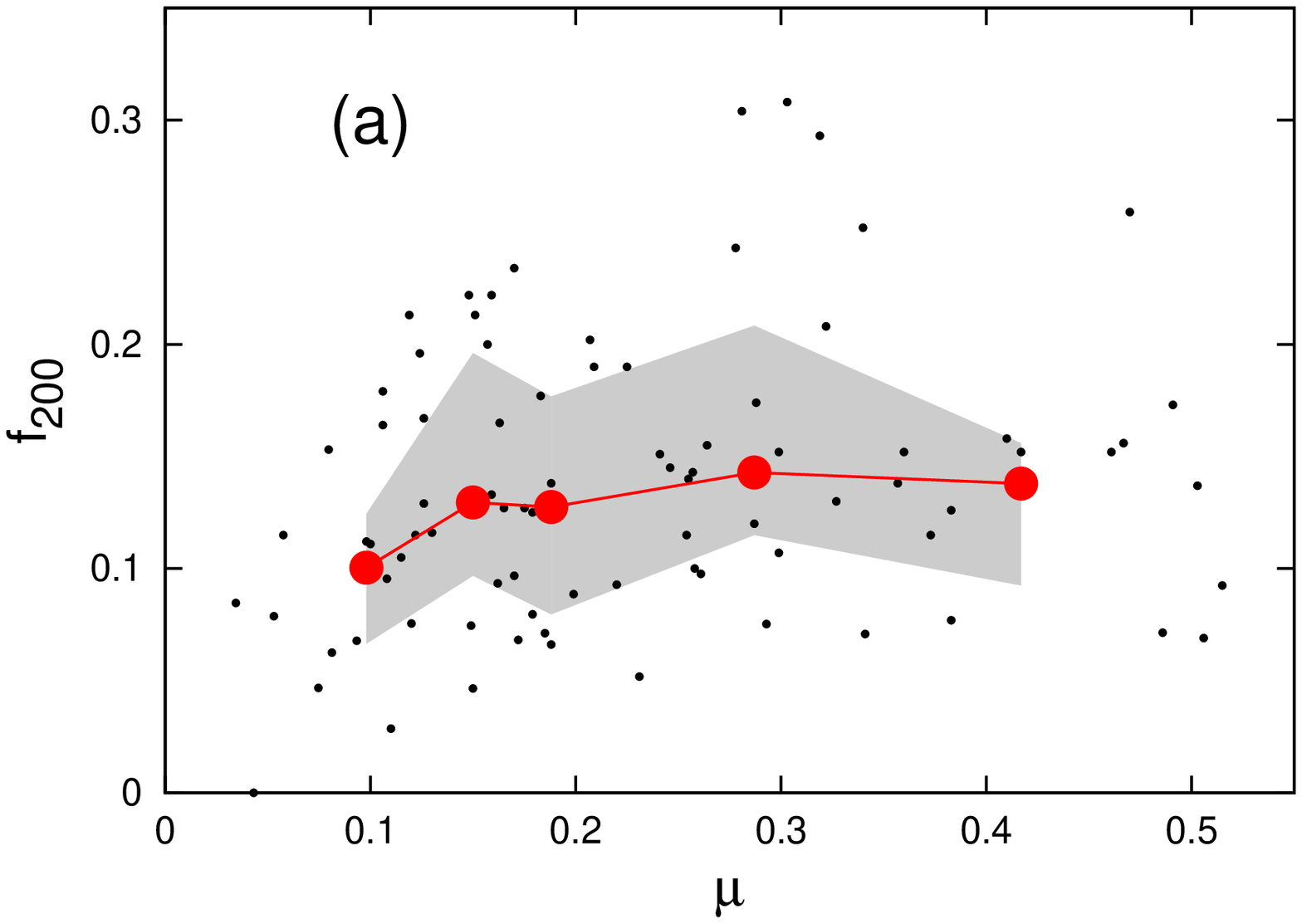} 
	\includegraphics[width=0.35\textwidth,clip,angle=270]{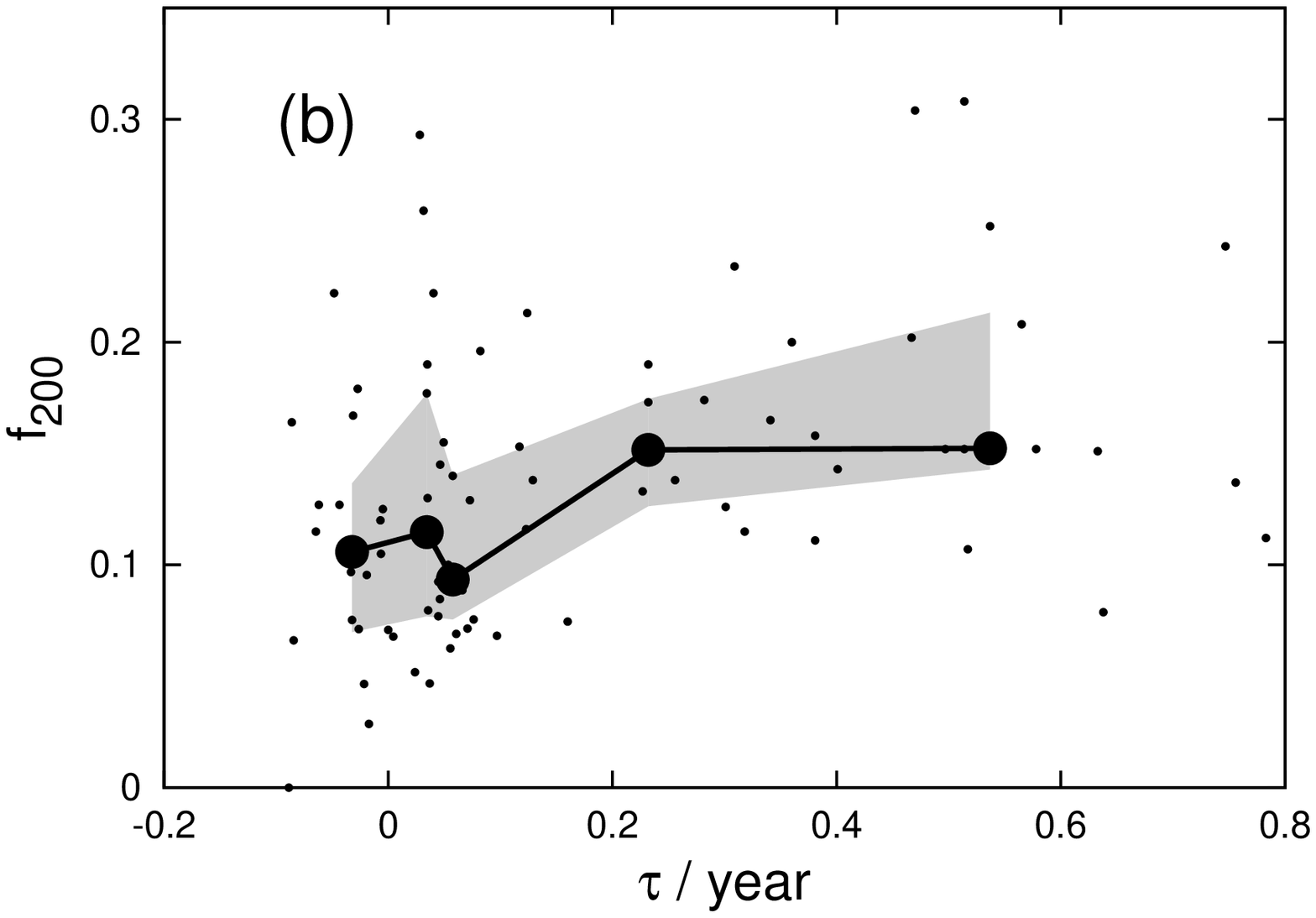} 
	\includegraphics[width=0.35\textwidth,clip,angle=270]{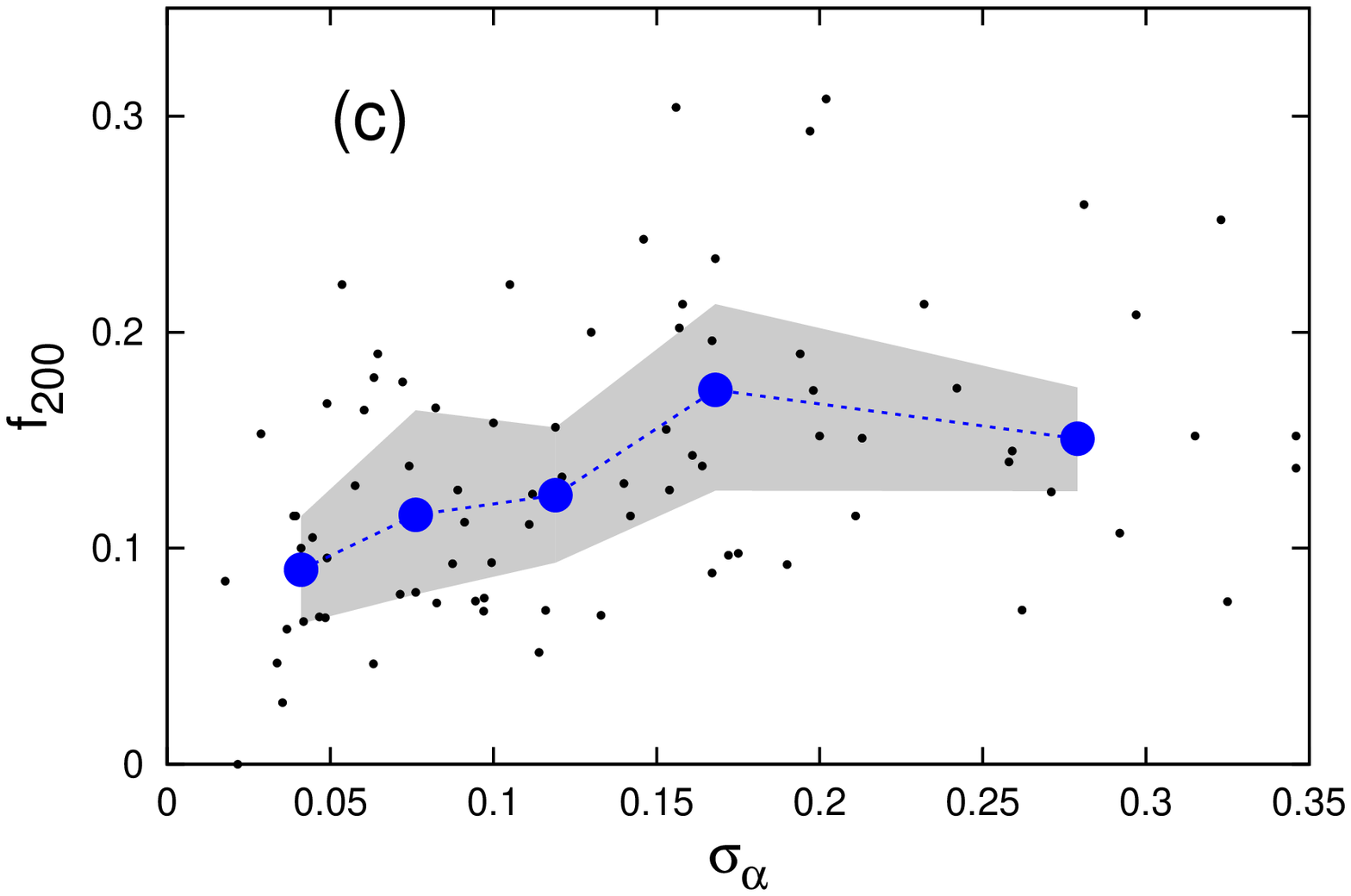} 
\caption{The dependence of source position stability on astrophysical indicators. {\it (a)}: X-band modulation index $\mu$; {\it (b)}: time lag $\tau$ between S and X-bands; {\it (c)}: spectral index rms $\sigma_{\alpha}$. In each case, the ordinate represents the fractions of positions $>200$~$\mu$as from the mean source position. Small black dots show all data; filled circles are bin medians. The data are binned to give an equal number of points in each bin. Shaded regions represent the 25 and 75\% confidence intervals.}
\label{fig:pos_vs_metrics}
\end{figure*}

Figure~\ref{fig:pos_vs_metrics} shows the dependence of source position stability on the three metrics for our sample of 84 segments. Each point represents a segment. In each panel, the ordinate shows $f_{200}$, the fraction of source positions that are further than 200~$\mu$as from the mean source position. The abscissa is X-band modulation index (top left panel), time lag between S and X-band light curves (top right panel) or spectral index rms (bottom panel). Filled circles and shaded regions show the same data divided into equal size bins; the circles are medians, and shaded regions the 25$^{\rm th}$ and 75$^{\rm th}$ percentiles.

Inspection of Figure~\ref{fig:pos_vs_metrics} shows that segments with small time lags and low spectral index variability appear to be associated with more stable positions (as indicated by lower values of $f_{200}$). This is borne out statistically: a standard Wilcoxon-Mann-Whitney test \citep{WallJenkins12} shows that segments with $\tau<0.06$~year (the sample median) have $f_{200}$ values that are significantly lower than segments with larger time lags; this difference is significant at the 99 percent level\footnote{We note that Monte Carlo simulations estimate that typical uncertainties on the derived time lags are $0.05$~year, and thus the $\tau<0.06$~year group is consistent with no measured time lag between S and X-band light curves.}. Segments with $\sigma_{\alpha}<0.12$ (sample median) also show more stable positions than those with higher values of $\sigma_{\alpha}$; this difference is also significant at the 99 percent level.

On the other hand, no clear relationship is seen between $f_{200}$ and the X-band modulation index $\mu$, except for a tendency to slightly lower $f_{200}$ values for the very lowest values of $\mu$ (the left-most bin in the top left panel of Figure~\ref{fig:pos_vs_metrics}). Segments with $\mu<0.13$ show marginally lower $f_{200}$ values (90 percent significance) than the rest of the sample, suggesting that positions of non-variable sources may be marginally more stable than their more variable counterparts.

Overall, our results suggest that low values of the time lag $\tau$ and spectral index rms $\sigma_{\alpha}$ are good indicators of source position stability. In this respect, it is important to note that the two variables are not independent: by construction, flaring sources with appreciable time lags will also exhibit significant spectral index variability. 

\section{Discussion}
\label{sec:discussion}

\subsection{Physical interpretation}
\label{sec:interpretation}


ICRF2 quasars are almost exclusively flat-spectrum radio sources. These are usually highly beamed, with the jet axis only a few degrees away from pointing directly at the observer \citep{PushkarevEA09}. As a result, high Doppler factors boost the radio emission significantly \citep{BlandfordKonigl79}; the emission will be boosted more strongly for jets aligned close to the observer's line of sight.

Synchrotron emission in the inner regions of flat-spectrum radio-loud quasars is highly self-absorbed, and to first order the bulk of emission at a given frequency comes from a relatively small region where the optical depth is close to unity, the so-called ``core'' \citep{Konigl81,Lobanov98,Hirotani05}. The observed core position{\footnote{The position referred to here is the astrometric position of the core as obtained from phase delays, for example via phase-referenced VLBI imaging. This is different to the geodetic source position derived from the group delay, which may or may not be affected by the core shift (see Section~\ref{sec:coreShift}).}} changes with frequency, moving closer to the base of the jet (the ``true'' core) for high frequencies. This is known as the core shift effect, and it has been measured through VLBI imaging \citep[e.g.][]{KovalevEA08,SokolovskyEA11}.

\begin{figure}
  	\includegraphics[width=0.5\textwidth,clip,angle=0]{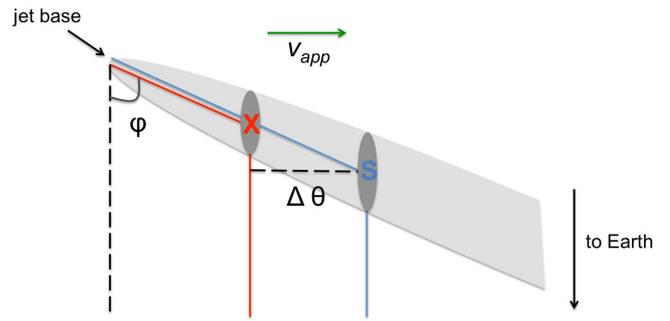} 
\caption{Time lag due to the core shift effect. The X-band ``core'' emission originates closer to the jet base than S-band ``core'' emission and therefore is observed first. The time lag depends on projected component separation $\Delta \theta$ and apparent velocity $v_{\rm app}$ of the jet component. Note that this apparent velocity can exceed the speed of light; this is known as superluminal motion \citep{Rees66} and is an artefact of projection effects and the finite speed of light.}
\label{fig:coreShift}
\end{figure}

The core shift effect will also manifests itself as a time lag between light curves at different frequencies. As we show in Figure~\ref{fig:coreShift}, a component co-located with the X-band core must move further down the jet before reaching peak emission at S-band. The relationship between projected core separation $\Delta \theta$, time lag $\tau$ and apparent component speed $v_{\rm app}$ is \citep[e.g.][]{KudryavtsevaEA11}

\begin{equation}
	\tau= D_{\rm A} \Delta \theta / v_{\rm app}
\label{eqn:tau}
\end{equation}
where the angular diameter distance $D_{\rm A}$ (derived from source redshift and an assumed cosmology) allows the observed angular separation of the two cores to be converted to projected physical separation. This is the origin of the time delay clearly seen in S and X-band light curves in Figure~\ref{fig:1144_lc}, characteristic of many ICRF2 sources.

From Figure~\ref{fig:coreShift} it is clear that, for a given surface brightness, jets that are {\t less} beamed (i.e. have larger values of $\varphi$) will exhibit more structure as viewed from Earth. They will also show greater core shifts (and as a consequence, larger time lags). This is consistent with our results in Figure~\ref{fig:pos_vs_metrics}b which show that sources with larger time lags appear to have less stable positions, an indication of more structure on larger scales. Time lag $\tau$ between the S and X-band light curves is therefore potentially a useful measure of projected source structure.

The intrinsic power of the outburst also has a role to play here: for a given value of $\varphi$, powerful outbursts will cause larger core shifts \citep[and therefore time lags;][]{SSG12}. This explanation also accounts for why there is only a weak dependence of position stability on the source modulation index (top panel of Figure~\ref{fig:pos_vs_metrics}). Highly beamed sources (i.e. sources with small values of $\varphi$) can still show large-amplitude flux density variations due to flaring; however they will not exhibit large time lags and will also show little projected structure. A testable prediction of this hypothesis is that highly beamed quasars lacking narrow emission lines (BL Lacs, blazars), should have smaller average core shifts than objects with narrow lines (quasars, Seyferts). Using the optical classifications of ICRF2 sources of \citet{MalkinTitov08}, we found no statistically significant difference between the two samples as given by the Kolmogorov-Smirnov test. We note that our analysis includes only 11 broad-line, and 33 narrow-line objects. Furthermore, effects such as detectability of the features (e.g. core shifts may be easier to detect in highly beamed sources) may also play a role. More rigorous tests with much larger, unbiased samples would be desirable to reliably confirm or refute this hypothesis.

\subsection{Reference frame improvement}


The results of Section~\ref{sec:results} suggest that the time lag between S and X-band light curves, or changes in the two point spectral index, may be used in deciding which sources should be scheduled for IVS observations. Accurate flux density monitoring at S and X-bands is required for this. How much improvement would be gained by scheduling sources in this manner? Table~\ref{tab:tau_medians} shows the position scatter weighted root mean square, medians and interquartile ranges for segments grouped by time lag $\tau$, spectral index rms $\sigma_\alpha$, and X-band modulation index $\mu$. Selecting sources with small time lags or low spectral index variability leads to a 20 percent improvement in astrometry, as measured by the source position weighted rms. In the longer term, a move to higher frequencies such as 24 and 43 GHz \citep{LanyiEA10,CharlotEA10} should reduce both source structure and core shifts.

\begin{table}
\caption{Distribution of source position scatter. Sources with time lags between S and X-bands that are less than the median value for the sample ($\tau<0.06$) have a position weighted rms that is 25 percent lower than sources with larger values of $\tau$. Similarly, sources with spectral index variability lower than the median ($\sigma_\alpha<0.12$) have positions that are 17 percent more stable than sources with higher values of $\sigma_\alpha$. No such difference is seen when sources are selected based on X-band modulation index.}
\label{tab:tau_medians}
\begin{tabular}{|c|cccc|}
\hline
Offset / $\mu$as & wrms & 1$^{\rm st}$ quartile & median & 3$^{\rm rd}$ quartile \\\hline\hline
$\tau<0.06$ & 13.5 & 4.5 & 8.2 & 15.7 \\
$\tau \geq 0.06$ & 16.9 & 5.1 & 9.5 & 18.1 \\\hline
$\sigma_{\alpha}<0.12$ & 14.0 & 4.6 & 8.3 & 15.8 \\
$\sigma_{\alpha} \geq 0.12$ & 16.3 & 4.9 & 9.3 & 17.9 \\\hline
$\mu<0.19$ & 15.5 & 4.8 & 8.8 & 17.1 \\
$\mu \geq 0.19$ & 15.1 & 4.7 & 8.8 & 16.7 \\
\hline
\end{tabular}
\end{table}

\subsection{Core shift, jet geometry, and implications for future reference frames}
\label{sec:coreShift}


Precise knowledge of the frequency dependence of quasar core positions has important implications for reference frame realisation. Writing the distance between the true jet base and the core observed at frequency $\nu$ as $r(\nu)=r_0 \nu^{-k_r}$, \citet{Porcas09} showed that the group delay due to the core shift is $t_{\rm cs, group} = (1 - k_r) r_0 \nu^{-k_r} L / c$, where $L$ is the projected baseline length and $c$ the speed of light. Thus, for $k_r=1$ the core shift group delay is exactly zero. The value of $k_r$ depends on jet geometry and the interaction between synchrotron emitting particles (which give rise to the radio emission) and magnetic field in the jet. For the simplest case of a conical jet (i.e. jet with a constant opening angle) with approximately equal energy densities in relativistic particles and magnetic field, $k_r$ is expected to be unity \citep{Konigl81}. \citet{KovalevEA08} and \citet{SokolovskyEA11} performed multi-frequency VLBI imaging of a number of radio-loud quasars, and found that $k_r \approx 1$ for sources in which they could reliably measure core shifts. On the other hand, \citet{KudryavtsevaEA11} used multi-frequency light curves to derive core shifts from time lags between different frequencies for the source 3C\,345. These authors found that the value of $k_r$ is significantly less than unity, and in fact depends on the amplitude of the flare being analysed. A similar result was recently found by \citet{MacquartEA13} for the quasar PKS~B1257--326.

\begin{figure}
  	\includegraphics[width=0.35\textwidth,clip,angle=270]{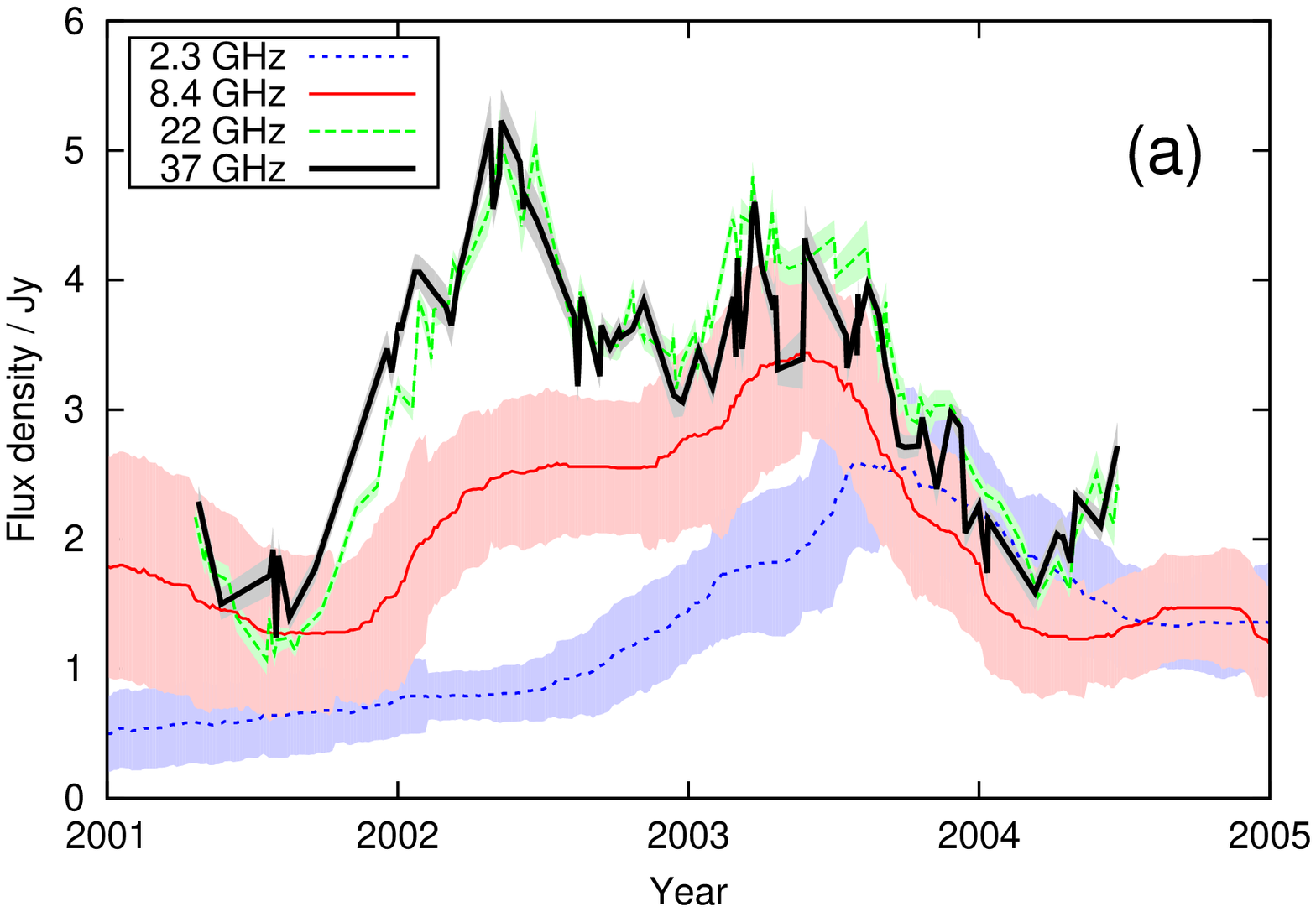} 
	\includegraphics[width=0.35\textwidth,clip,angle=270]{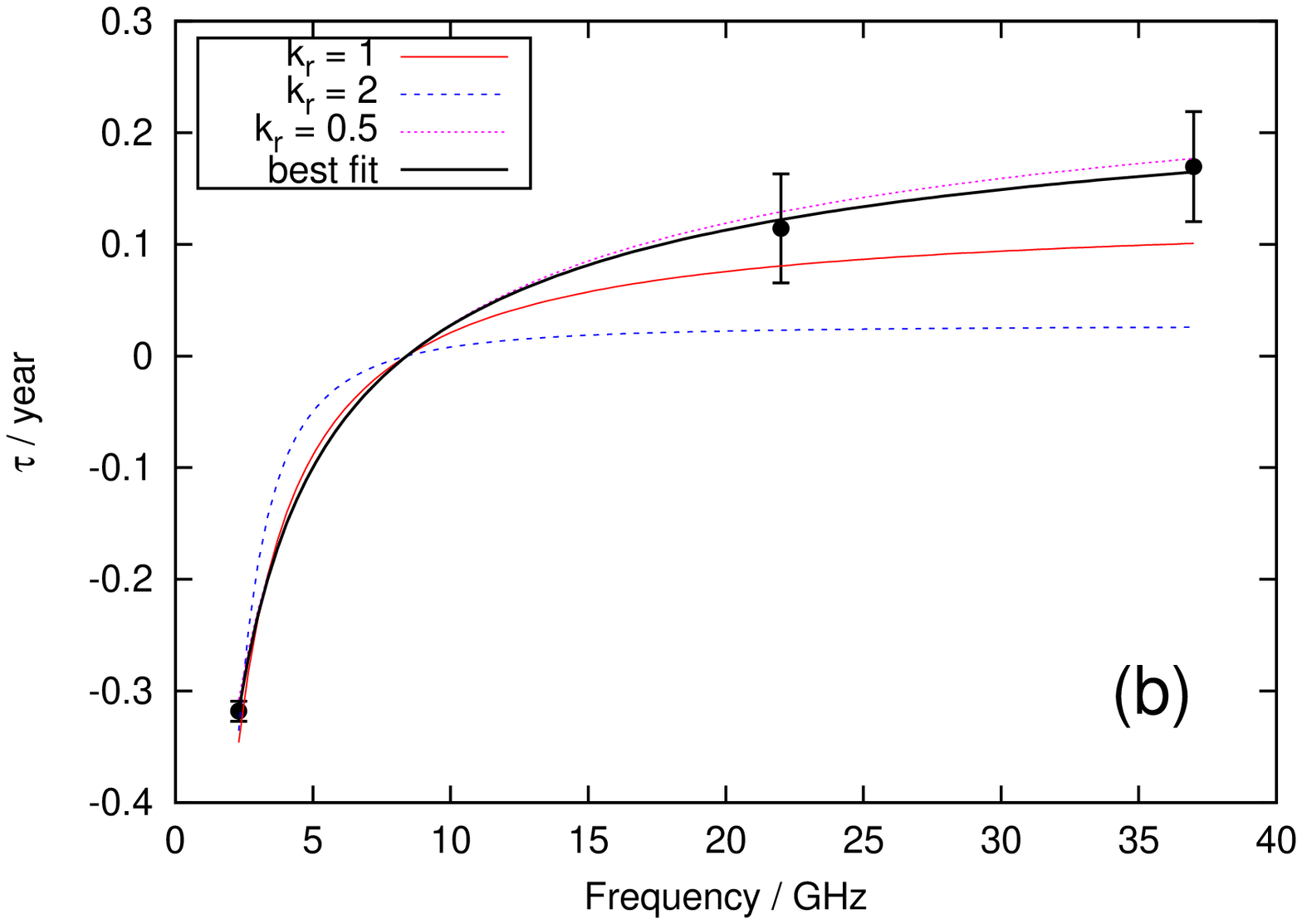} 
\caption{Multi-frequency analysis for a flare in source B0059+581. {\it (a)}: light curves. The data are from IVS (S and X-bands), and \citet{TerasrantaEA05}(22 and 37 GHz). {\it (b)}: time lags as a function of frequency separation. The reference frequency is 8.4~GHz. Uncertainties were calculated through DCF analysis of Monte-Carlo simulations of light curves at each frequency. Fits for various $k_r$ values are shown. The data are inconsistent with $k_r=1$, the expected value for a non-accelerating conical jet at equipartition which yields zero group delay core shift.}
\label{fig:0059}
\end{figure}

In Figure~\ref{fig:0059} we perform a similar analysis for the ICRF2 quasar B0059+581. We use IVS flux density time series at 2.3 and 8.4~GHz together with data from the Mets\"ahovi observatory single-dish monitoring program at 22 and 37~GHz \citep{TerasrantaEA05}. We use the DCF to determine time lags between the light curves in the top panel of Figure~\ref{fig:0059}. The three independent frequency pairs are plotted in the bottom panel. We fit various values of $k_r$ by using the time lags between frequencies $\nu_1$ and $\nu_2$ \citep[e.g.][]{KudryavtsevaEA11},

\begin{equation}
\Delta t_{1,2} \propto \left( \frac{\nu_2}{\nu_{\rm ref}} \right)^{-k_r} - \left( \frac{\nu_1}{\nu_{\rm ref}} \right)^{-k_r}
\label{eqn:delta_t}
\end{equation}
where $\nu_{\rm ref}$ is a reference frequency ($\nu_{\rm ref}=8.4$~GHz in Figure~\ref{fig:0059})\footnote{Equation~\ref{eqn:delta_t} can also be used to validate the use in Section~\ref{sec:fluxData} of 6.7~GHz data as a proxy for 8.4~GHz variability: for $k_r=0.6$, the time lag between 6.7 and 8.4 GHz is less than 15 percent of the time lag between 2.3 and 8.4 GHz; and this value is even smaller for larger $k_r$.}. We find a best fit of $k_r=0.58 \pm 0.04$. Hence our data are inconsistent with $k_r=1$.

There is no inconsistency between our findings \citep[and those of][]{KudryavtsevaEA11} of $k_r<1$ and core shift measurements obtained through VLBI imaging. First of all, the two methods are sampling very different physical scales. Dedicated monitoring of the quasar B0059+581 by the MOJAVE team has shown that new jet components move at apparent angular speeds of $0.2-0.3$~mas~year$^{-1}$ \citep{ListerEA09b}. A time lag of 0.32 years detected between S and X bands for B0059+581 can therefore be converted to a projected separation of $70-90$~$\mu$as. By contrast, the median detected core shift between these same frequencies for the sample of \citet{KovalevEA08} is 440~$\mu$as. We note that the core shifts measured through VLBI imaging suffer strongly from selection bias: \citet{KovalevEA08} reliably measured core shifts in only 29 of 277 targeted radio sources. It is thus entirely possible that a large population of sources with core shifts much less than $100$~$\mu$as exists. It has been suggested by \citet{Lobanov98} on theoretical grounds that quasar jets should have $k_r<1$ on small scales, and gradually tend to $k_r=1$ as equilibrium is established between the energy densities of relativistic particles and magnetic field. The other possible explanation is component acceleration: if jet components do not move at a constant apparent velocity, the time lag in Equation~\ref{eqn:tau} is no longer linearly related to the core shift.

While this result is very interesting astrophysically, it also has important implications for astrometry and geodesy. Knowledge of core position as a function of frequency is necessary for broadband delay measurements planned for next-generation VLBI systems \citep{PetrachenkoEA09}. They are also vital for accurate alignment of the radio reference frame with the Gaia optical frame \citep{BourdaEA10,BourdaEA11}.

Multi-frequency flux density monitoring has an important role to play even in current geodetic VLBI observations. Multi-frequency light curve analysis allows us to probe angular scales that are almost an order of magnitude smaller than conventional VLBI imaging. Because these observations do not require a large number of antennas, they can be performed at much higher cadence than VLBI imaging. The use of flux density data potentially allows us to circumvent the so-called sub-resolution problem, and estimate the amount of source structure on sub-100~$\mu$as scales that are important for geodesy.

\section{Conclusions}
\label{sec:conclusions}


We used International VLBI Service for Geodesy and Astrometry observations to study the flux density variability of ICRF2 quasars. Construction and analysis of S and X-band light curves allows us to probe the amount of quasar structure on sub-100$~\mu$as scales through correlation with source position stability. This method is complementary to lower-cadence conventional VLBI imaging studies.

Based on physical models of quasars, we constructed three metrics of quasar variability: {\it (a)} modulation index (a measure of variability in source flux density); {\it (b)} time lag between the S and X-band light curves; and {\it (c)} spectral index variability (how much the ratio of X-to-S band flux density changes over time). We tested whether these metrics correlate with source position stability as determined from standard geodetic VLBI solutions.

We found that the time lag and spectral index rms are good indicators of position stability. Selecting sources with small time lags and/or spectral index rms reduces the weighted rms in source positions by 20 percent. This suggests that multi-frequency light curves provide an important way of probing the size of small-scale VLBI structure.

We illustrate that multi-frequency light curves allow us to study in detail the effects of the frequency-dependent core shift. Using the example of quasar B0059+581, we find that the core shift dependence on frequency may depart significantly from the $\nu^{-1}$ dependence expected for the simplest case of a conical jet at equipartition on sub-100~$\mu$as scales. These findings have important implications for both geodetic VLBI, and future alignment of the radio and optical reference frames.

\begin{acknowledgements}

SS and JM thank the Australian Research Council for Super Science Fellowships (FS100100037 and FS110200045). We are grateful to Harald Schuh, Simon Ellingsen and John Dickey for useful discussions, and Chris Jacobs and three other anonymous referees for thorough and constructive comments that have significantly improved the manuscript. This research has made use of the United States Naval Observatory (USNO) Radio Reference Frame Image Database (RRFID).

\end{acknowledgements}

\bibliographystyle{apj}
\bibliography{apj-jour,TwoPointSpecIndex}   

\end{document}